\documentclass[a4paper,11pt]{article}
\usepackage{amsfonts}
\usepackage{amsmath}
\usepackage{amssymb}
\usepackage{graphicx}
\usepackage{array}
\usepackage{booktabs}
\usepackage{multirow}
\usepackage{makecell}
\usepackage{float}
\pdfoutput=1 

\usepackage{mymacros}
\usepackage{chngcntr}
\usepackage{verbatim}
\usepackage{amsthm}
\usepackage{graphics}
\usepackage{xcolor}
\usepackage{mathrsfs}
\usepackage{bbold}
\usepackage{cases}
\usepackage{epsfig}
\usepackage{epstopdf}
\usepackage{hyperref}
\usepackage{amsfonts}
\usepackage{hyperref}
\usepackage{jheppub} 

\usepackage[T1]{fontenc} 

\hypersetup{
	colorlinks=true,
	linkcolor=blue,
	filecolor=blue,
	urlcolor=blue,
	citecolor=blue,
}

\title{\boldmath Angular momentum and chaos bound of charged particles around Einstein-Euler-Heisenberg AdS black holes}

\author[a,1]{Deyou Chen,}
\author[a,2]{Chuanhong Gao}

\emailAdd{deyouchen@hotmail.com}
\emailAdd{chuanhonggao@hotmail.com}

\affiliation{$^{a}$School of Science, Xihua University, Chengdu 610039, China}

\abstract{In this paper, we investigate influences of the charge and angular momentum of a particle around a charged Einstein-Euler-Heisenberg AdS black hole on a Lyapunov exponent, and find spatial regions where the chaos bound is violated. Positions of circular orbits are gotten by fixing the charge and angular momentum of the particle, respectively. The positions gradually move away from the event horizon with the increase of the angular momentum when the charge is fixed and with the decrease of the charge when the angular momentum is fixed. For certain values of the charge, angular momentum and Euler-Heisenberg parameter, the spatial regions where the bound is violated are found. When the charge is fixed and the Euler-Heisenberg parameter is large, a small angular momentum causes the violation. Although the angular momentum is small, the corresponding spatial region is not small. An interesting discovery is that the bound is violated by the black hole when the particle’s charge is less than 1  and $\Lambda =0$, but this requires the black hole's charge to be large enough. This violation may be related to the dynamical stability of the black hole. The backreaction of the particle on the background spacetime isn't considered in the investigation.}


\begin{document} 
\maketitle
\flushbottom

\section{Introduction}

Motions of particles around black holes, especially in their equatorial planes, attract much attention. The reason is that these motions convey important information on the background spacetimes. For example, null geodesics are useful to explain quasinormal modes (QNMs) of a black hole for the test fields \cite{CMBWZ,KZS1,KZS2}. The real part of QNMs is determined by the angular velocity of a particle at the unstable orbit, and the imaginary part is related to the unstable timescale of the orbit. Unstable circular geodesics affect the optical appearance of a gravitationally collapsed star, and then can explain the star's luminosity \cite{AT}. Moreover, the spins and merger of black holes can also be explained by circular motions of particles \cite{PK}. 

The motions of particles may cause chaos. Chaos is an important phenomenon in nonlinearly dynamic systems. In a classical chaos, a particle’s trajectory deviates exponentially from it's initial location due to a small change of initial conditions. In quantum systems, an important probe of chaotic behaviors is out-of-time ordered correlators (OTOCs) \cite{HS1,HS2,RRP,JKY,LR}. It was first computed by Shenker and Stanford using the eikonal approximation. They found that the Lyapunov exponent is related to the temperature of the black hole \cite{HS2}. 
Recently, Maldacena, Shenker and Stanford conjectured that there is a universal upper bound for the exponent of the chaos in the thermal quantum systems with a large number of degrees of freedom,

\begin{eqnarray}
\lambda \leq \frac{2\pi T}{\hbar},
\label{eq1.1}
\end{eqnarray}

\noindent where $\lambda$ denotes the Lyapunov exponent, and $T$ is the temperature of the system \cite{MSS}. This conjecture plays an important role in black hole physics and gravity duality. After this seminal work was put forward, it immediately attracted much attention and was verified by a large number of work \cite{PR,MS,CFCZZ,HT,HT1,HT2,KS,HG,BHM,FK,HRY,CSSTW,BPK,LLW,WHH,GSS,CRYZ,LX,CJLY,WWT,CDP,CDP2,NNT,CLC,EA,JMW,CCB,BCR,GJT,HSBV,EFST,BD,DMM2,ACO,LG1,MPD,GR,CCKM1,BL,MRS,LTW1,LTW2,PP1,PP2,PP3,CWJ1,MWZ1,LW,CKR1,CKR2,HY,CFL,GZK1,GZK2,ADNT,ADNT1,ADNT2,ADNT3,QCL}. The study of the chaos in the Sachdev-Ye-Kitaev (SYK) model is a successful  verification. It was found that the SYK model saturates the chaos bound, which leads to a speculation that the saturation of the bound is a necessary condition for Einstein gravity dual. For a single-particle system, considering that a particle moves radially and is close to a black hole by a strong enough external forces without falling into it, Hashimoto and Tanahashi found that the exponent is not related to the species and strength of the forces, but only to the black hole's surface gravity, namely $\lambda \leq \kappa$, where $\kappa$ is the surface gravity \cite{HT}. From the relation between the surface gravity and temperature, this result is consistent with the conjecture in \cite{MSS}. 

However, there are cases that the chaos bound is violated. When a probe particle moves around a charged black hole, the particle can be in equilibrium in a circular orbit by centrifugal or Lorentz forces. For a neutral particle, a force on it is a centrifugal force, and this force is related to the particle's angular momentum. The angular momentum causes the change of the effective potential, but it is not sufficient to make the particle’s orbit closer to the event horizon. For a charged particle, one can adjust its charge-to-mass ratio to make it close to the horizon. Based on this, Zhao et al expanded the exponent at the horizon and studied the chaos bound of the charged particles near the horizons of the charged black holes \cite{ZLL}. They found that the bound was satisfied by the Reissner-Nordstr\"om (RN) and Reissner-Nordstr\"om anti-de Sitter (RN AdS) black holes, and violated by a large number of black holes. They obtained the exponent by using the effective potential method. The contribution of the angular momenta of the particles was neglected. In fact, the angular momenta not only affect the effective potentials, but also increases the magnitudes of the chaotic behaviors. When this influence was considered, the auhors discussed the bound of the charged particles around the rotating charged black holes, and found that the bound is violated \cite{KG1,KG2}. In this work, the exponent was also obtained by the effective potentials. In \cite{LG2}, Lei and Ge got the exponent by solving the eigenvalue of the Jacobian matrix. The bound in the RN and RN AdS black holes was studied by the expansions of the exponent at the horizon. When the angular momentum's influence is taken into account, the bound is violated.   

In this paper, we investigate influences of the charge and angular momentum of a particle around a charged Einstein-Euler-Heisenberg (EEH) AdS black hole on a Lyapunov exponent, and find spatial regions where the chaos bound is violated. The influence of the Euler-Heisenberg (EH) parameter on the exponent is discussed in detail. The exponent is obtained by solving a determination of eigenvalues of a Jacobian matrix in a phase space. The particle's charge is fixed to obtain the exponent's expression in our calculation, which is different from that in \cite{LG2}. At the same time, we don't expand the exponent at the event horizon. Based on this derivation, we numerically discuss the exponent by fixing the charge and angular momentum, respectively. The angular momentum plays an important role in the investigation. It affects not only the value of the exponent, but also positions of circular orbits. We find the positions and spatial regions where the bound is violated by fixing the particle's charge and changing its angular momentum. 

Nonlinear electrodynamics (NED) is a generalization of Maxwell’s theory  and can be viewed as a low-energy limit from string theory or D-brane physics, where Abelian and non-Abelian NED Lagrangians could be produced \cite{YCLW}. In \cite{BI}, Born and Infeld put forward a theory where the central singularity of the electromagnetic field of a point charge could be removed to study the NED. Based on this theory, some interesting black hole solutions were gotten and their properties were studied. In \cite{HE}, Euler and Heisenberg first constructed the Lagrangian of the NED by using the Dirac electron-positron theory. This Lagrangian has the high-order terms of the nonlinear electromagnetic field, and was reconstructed by Schwinger in the frame of quantum electrodynamics (QED) \cite{JSS1}. The reconstructed Lagrangian carries the main features of the EH NED and characterizes the phenomenon of vacuum polarization. If the strength of an electric field is higher than the critical value ($m^2c^3/e\hbar $), the QED effect leads to the emergence of particle pairs in the vacuum. A more general formulation of the NED which admits an arbitrary function of the electromagnetic invariants was later developed in \cite{JP1}. Based on the four-dimensional action of general relativity coupled to the EH NED$^{1}$ \cite{JP2,SGP}\note{A short review of the EH effective Lagrangian was given in \cite{RWX2}. A summary of the EH NED and the static spherically symmetric solution of the EEH equations were given in \cite{RWX5}.}, some black hole solutions with or without the cosmological constant have been obtained \cite{RWX1,RWX2,RWX3,RWX4,AM,MB,RWX5,RWX6}. The EEH AdS black hole is one of them. In addition, the EH action is approximated to the supersymmetric system of the minimum coupling spin-1/2, -0 particles with appropriate parameters. When the field has high intensity, this is a more accurate QED classical approximation than Maxwell's theory \cite{RWX1}. QED provides a solid foundation for the electromagnetic interaction. Moreover, it has been experimentally verified \cite{RVX}. Therefore, it is very important to study QED effects in black hole physics. On the other hand, an important feature of AdS black holes is the existence of the phase transitions and critical phenomenon, where the cosmological constant is regarded as a thermodynamic pressure and its conjugate quantity is a thermodynamic volume. Recent developments in the AdS/CFT correspondence greatly stimulate the study of the AdS black holes. 

The rest is organized as follows. In the next section, we obtain the Lyapunov exponent by solving the eigenvalues of the Jacobian matrix in a phase space $(\pi_r, r)$. In Section 3, the positions of circular orbits of the particle around the EEH AdS black hole are obtained. The values of the exponent are calculated by considering the influences of the angular momentum and charge, and the spatial regions where the bound is violated are found. The last section is devoted to our conclusions and discussions.

\section{Lyapunov exponent in EEH AdS black holes}

The four-dimensional action of general relativity with the cosmological constant coupled to the NED is  \cite{JP2,SGP}

\begin{eqnarray}
S=\frac{1}{4\pi}\int_{M^4}d^4x\sqrt{-g}\left[\frac{1}{4}(R-2\Lambda) - \mathcal{L}(F,G)\right],
\label{eq2.1}
\end{eqnarray}

\noindent where $R$ is the Ricci scalar, and $\Lambda$ is the cosmological constant. $\mathcal{L}(F,G)$ is the NED Lagrangian which depends on the electromagnetic invariants, $F=\frac{1}{4}F_{\mu\nu}F^{\mu\nu}$, $G=\frac{1}{4}F_{\mu\nu}$$^{\star} F^{\mu\nu} $, $F_{\mu\nu}$ is the electromagnetic field strength tensor and its dual is $^{\star} F^{\mu\nu} =\frac{\epsilon_{\mu\nu\sigma\rho}F^{\sigma\rho}}{2\sqrt{-g})} $. $\epsilon_{\mu\nu\sigma\rho}$ is the antisymmetric tensor and satisfies $\epsilon_{\mu\nu\sigma\rho}\epsilon^{\mu\nu\sigma\rho} =-4!$. The Lagrangian density for the EH NED is \cite{HE}

\begin{eqnarray}
\mathcal{L}(F,G)=- F+\frac{a}{2}F^2 +\frac{7a}{8}G^2 ,
\label{eq2.1.1}
\end{eqnarray}

\noindent where $a=\frac{8\alpha^2}{45m^4}$ is the EH parameter, $\alpha$ is the fine structure constant and $m$ is the electron mass. In \cite{MB}, Magos and Bretón got a static spherically symmetric solution from the action (\ref{eq2.1}). The solution is given by 

\begin{eqnarray}
ds^2 = -F(r)dt^2 + \frac{1}{F(r)}dr^2 + r^2(d\theta^2+\sin^2\theta d\phi^2),
\label{eq2.2}
\end{eqnarray}

\noindent with an electromagnetic potential 

\begin{eqnarray}
A_t=\frac{Q}{r}-\frac{aQ^{3}}{10r^{5}},
\label{eq2.3}
\end{eqnarray}

\noindent where

\begin{eqnarray}
F(r)=1-\frac{2M}{r}+\frac{Q^{2}}{r^{2}}-\frac{\Lambda r^{2}}{3}-\frac{aQ^{4}}{20r^{6}},
\label{eq2.4}
\end{eqnarray}

\noindent $M$ and $Q$ are the mass and electric charge of the black hole, respectively. The cosmological constant $\Lambda$ can be positive or negative values. In this paper, we focus our attention on the AdS spacetime and then the constant is negative. When $\Lambda = 0$, the metric describes a charged EEH black hole gotten by Amaro and Macias \cite{AM}. When $a = 0$, the metric describes RN AdS black holes. Due to the existence of the EH parameter, there is an extra term compared to the metric of the RN AdS black holes and its effect is of reinforcing the gravitational attraction. Therefore, the behavior of this system is more like Schwarzschild AdS black holes than RN AdS black holes. There is the opposite sign of the charge terms in the solution, and the minus sign makes the divergences of the solution opposite to those of the RN AdS black holes. The term $\frac{Q^2}{r^2}$ is always bigger than $\frac{a Q^4}{20r^6}$ outside the event horizon when $0\leq a \leq \frac{32Q^2}{7}$. This condition ensure that the dominant behavior of the solution is given by the linear theory. The event horizon $r_+$ is determined by $F(r) = 0$. The surface gravity is 

\begin{eqnarray}
\kappa=\frac{1}{2r_+}\left(1 - \frac{Q^2}{r_+^2}-\Lambda r_+^2+\frac{aQ^4}{4r_+^6}\right).
\label{eq2.5}
\end{eqnarray}

A Lyapunov exponent denotes the average rate of contraction and expansion of two adjacent orbits in a classical phase space. It can be derived by the effective potential method and the matrix method. Here we review the Lyapunov exponent which has been obtained by the matrix method in \cite{CMBWZ,PP1,PP2,PP3,LG1,LG2}. From the equations of motion of a particle, 

\begin{eqnarray}
\frac{dX_i(t)}{dt} = F_i(X^j),
\label{eq2.1.2}
\end{eqnarray}

\noindent we use a certain orbit and linearize the equations,

\begin{eqnarray}
\frac{d\delta X_i(t)}{dt} =K_{ij}(t)\delta X_j(t),
\label{eq2.1.3}
\end{eqnarray}

\noindent where $K_{ij}(t) $ is a linear stability matrix defined by $K_{ij}(t) = \left.\frac{\partial F_i}{\partial X_j}\right|_{X_i(t)}$. The solution of Eq. (\ref{eq2.1.3}) satisfies 

\begin{eqnarray}
\delta X_i(t) = L_{ij}(t)\delta X_j(0),
\label{eq2.145}
\end{eqnarray}

\noindent where $L_{ij}(t)$ is a evolution matrix and obeys $ \dot{L}_{ij}(t)=K_{im} L_{mj}(t)$ and $L_{ij}(0) = \delta_{ij}$. The exponent is obtained from the determination of the eigenvalues of the matrix $L_{ij}$. It can be written as 

\begin{eqnarray}
\lambda = \lim_{t \to \infty} \frac{1}{t}ln\left(\frac{L_{jj}(t)}{L_{jj}(0)}\right).
\label{eq2.1.5}
\end{eqnarray}

\noindent The positive exponent implies a chaos in the system. When we focus our attention on a circular orbit in an equatorial plane of a black hole, the classical phase space is expressed as $(p_r, r)$. Let's further linearize the equation of motion with $X_i(t) = (p_r, r)$ about the orbit of constant $r$, and then get $K_{11}=K_{22}=0$,  $K_{12}=\frac{d}{dr}\left(\dot{t}^{-1}\frac{\delta \mathcal{L}}{\delta r}\right)$ and $K_{21}=-(\dot{t}g_{rr})^{-1}$, where $\mathcal{L}$ is the Lagrangian for geodesic motion. Thus, the exponent is gotten as

\begin{eqnarray}
\lambda = \sqrt{K_{12}K_{21}}.
\label{eq2.1.6}
\end{eqnarray}

When a test particle with charge $q$ moves in a circle around the equator of the black hole, its Lagrangian is \cite{CMBWZ,PP1,PP2,LG1,LG2,WL1}

\begin{eqnarray}
\mathcal{L} = \frac{1}{2}\left(-F\dot{t}^2+\frac{\dot{r}^2}{F} +r^2\dot{\phi}^2\right) -qA_t\dot{t},
\label{eq2.6}
\end{eqnarray}

\noindent where $\dot{x^{\mu}} = \frac{dx^{\mu}}{d\tau}$ and $\tau$ is a proper time. From the definition of the generalized momentum $ \pi_{\mu}=\frac{\partial\mathcal{L}}{\partial\dot{x}}$, we get

\begin{eqnarray}
\pi_t = -F\dot{t} -qA_t=-E, \quad\quad \pi_r = \frac{\dot{r}}{F}, \quad\quad \pi_{\phi} = r^2 \dot{\phi}=L.
\label{eq2.7}
\end{eqnarray}

\noindent In the above derivation, $E$ is the energy of the particle, and $L$ denotes the angular momentum. Then the  Hamiltonian is

\begin{eqnarray}
H = \frac{-(\pi_{t}+qA_{t})^2+\pi_r^2F^2+ \pi^2_{\phi}r^{-2}F}{2F}.
\label{eq2.8}
\end{eqnarray}

\noindent The motion equations of the particle are gotten from the Hamiltonian, which are

\begin{eqnarray}
\dot{t} &=& \frac{\partial H}{\partial \pi_t}=-\frac{\pi_t+qA_t}{F}, \quad  \dot{\pi_t}= -\frac{\partial H}{\partial t} =0 ,
\quad \dot{r} = \frac{\partial H}{\partial \pi_r}= \pi_rF, \nonumber\\
\dot{\pi_r} &=& -\frac{\partial H}{\partial r} =-\frac{1}{2}\left[\pi^2_r F^{\prime} -\frac{2qA^{\prime}_t(\pi_t+qA_t)}{F}+\frac{(\pi_t+qA_t)^2F^{\prime}}{F^2} -\pi^2_{\phi}(r^{-2})^{\prime}\right], \nonumber\\
\dot{\phi} &=& \frac{\partial H}{\partial \pi_{\phi}}= \frac{\pi_{\phi}}{r^2}, \quad  \dot{\pi_\phi}= -\frac{\partial H}{\partial \phi} =0.
\label{eq2.9}
\end{eqnarray}

\noindent In the above equations, "$\prime$" denotes the derivative with respect to $r$. From the equations of motion, the relations between the radial coordinate and time, and between the radial momentum and time are obtained.

\begin{eqnarray}
\frac{dr}{dt} &=& \frac{\dot{r}}{\dot{t}} =-\frac{\pi_rF^2}{\pi_t+qA_t}, \nonumber\\
\frac{d\pi_r}{dt} &=& \frac{\dot{\pi_r}}{\dot{t}} = -qA^{\prime}_t +\frac{1}{2}\left[\frac{\pi^2_r FF^{\prime}}{\pi_t+qA_t}+\frac{(\pi_t+qA_t)F^{\prime}}{F} -\frac{\pi^2_{\phi}(r^{-2})^{\prime}F}{\pi_t+qA_t}\right].
\label{eq2.10}
\end{eqnarray}

\noindent For convenience, we define $F_1= \frac{dr}{dt}$ and $F_2=\frac{d\pi_r}{dt}$. The normalization of the four-velocity of a particle is given by $ g_{\mu\nu}\dot{x}^{\mu}\dot{x}^{\nu}=\eta$, where $\eta =0$ describes a case of a massless particle and $\eta =-1$ describes a case of a massive particle. A charged particle is considered, and then the normalization yields a constrain condition, $\pi_t+qA_t=-\sqrt{F(1+ \pi_r^2F + \pi_{\phi}^2r^{-2})}$. We use this constrain and rewrite Eq. (\ref{eq2.10}) as

\begin{eqnarray}
F_1 &=& \frac{\pi_rF^2}{\sqrt{F(1+ \pi_r^2F + \pi_{\phi}^2r^{-2})}}, \nonumber\\
F_2 &=& -qA^{\prime}_t -\frac{(2\pi^2_r F+1)F^{\prime}}{2\sqrt{F(1+ \pi_r^2F + \pi_{\phi}^2r^{-2})}} -\frac{\pi^2_{\phi}(r^{-2}F)^{\prime}}{2\sqrt{F(1+ \pi_r^2F + \pi_{\phi}^2r^{-2})}}.
\label{eq2.11}
\end{eqnarray}

The effective potential of the particle plays an important role in the acquisition of the exponent \cite{ZLL,KG1,KG2}. Here the exponent is derived by the eigenvalues of a Jacobian matrix in the phase space $(r, \pi_r)$. In this phase space, the Jacobian matrix is defined by $K_{ij}$ with the elements 

\begin{eqnarray}
K_{11} = \frac{\partial F_1}{\partial r} , \quad\quad
K_{12} = \frac{\partial F_1}{\partial \pi_r} ,\quad\quad
K_{21} = \frac{\partial F_2}{\partial r} ,\quad\quad
K_{22} = \frac{\partial F_2}{\partial \pi_r}.
\label{eq2.12}
\end{eqnarray}

\noindent When the particle is in equilibrium, its trajectory satisfies

\begin{eqnarray}
\pi_r=\frac{d\pi_r}{dt} =0.
\label{eq2.13}
\end{eqnarray}

\noindent Then the exponent of the chaos of the charged particle in an circular orbit is obtained by calculating the eigenvalues of the matrix, which is

\begin{eqnarray}
\lambda^2 = \frac{1}{4}\left[\frac{F^{\prime}+\pi_{\phi}^2(r^{-2}F)^{\prime}}{1+\pi_{\phi}^2r^{-2}} \right]^2 -\frac{1}{2}F\frac{F^{\prime\prime}+\pi_{\phi}^2(r^{-2}F)^{\prime\prime}}{1+\pi_{\phi}^2r^{-2}}
-\frac{qA_t^{\prime\prime}F^2}{\sqrt{F(1+\pi_{\phi}^2r^{-2})}}.
\label{eq2.14}
\end{eqnarray}

\noindent At the event horizon, $F(r_+)=0$. We get 

\begin{eqnarray}
\lambda^2 =\frac{1}{4}F^{\prime}=\kappa^2,
\label{eq2.15}
\end{eqnarray}

\noindent which is saturated at the horizon. This result is consistent with that obtained in the spherically symmetric black holes \cite{ZLL,LG2}. In the derivation of Eq. (\ref{eq2.14}), to find the range of the angular momentum and spatial region where the bound is violated, we fixed the particle’s charge in the calculation process, and didn't expand the exponent at the event horizon. When the charge is expressed in terms of $\pi_{\phi}$, $r$ and $A_t$, the expression of the exponent in \cite{LG2} is easily recovered. It is clearly that the exponent is affected by the particle's angular momentum and charge. When the angular momentum is neglected and the expression of the charge is adopted, the exponent obtained in \cite{ZLL} is recovered. In the following, we investigate the influences of the charge and angular momentum on the exponent, and find spatial regions where the bound is violated. 

\section{Chaos bound and its violation by EEH AdS black holes}

When a charged particle moves around a charged black hole, the position of the particle's circular orbit is affected by its charge and angular momentum. One can let the particle infinitely close to the horizon by adjusting the charge-to-mass ratio and angular momentum. The chaos bound in the near-horizon regions has been studied by the expansions of the Lyapunov exponent at the horizon. Here we investigate the influences of the angular momentum and charge on the exponent and find spatial regions where the bound is violated. The region we are concerned about is not limited to the near-horizon region, but also at a certain distance from the horizon. The cosmological constant is a very small quantity in physics. In some papers, people often took a small value of the constant to study various problems \cite{WL1}. We also found that the constant's value in some papers is relatively large, such as \cite{LG2,KG2,MB}, etc. In \cite{KG2}, the authors ordered the constant to be $-0.5$ and $-1$ for the RN AdS and Kerr-Newman AdS black holes.  In \cite{LG2}, the authors ordered the black holes’ charge to be $1$, and the constant to be $-3$ for the RN AdS black holes. In \cite{MB}, the black holes’ mass is $1$, the charge is $0.8$, and the constant is $-3$ for the RN AdS and EEH AdS black holes. In view of these values of the constant, we order $\Lambda=- 3$ and $\Lambda=0$ in this section, respectively.

We first find the positions of the orbits by numerical calculations. When $\Lambda=-3$, Eq. (\ref{eq2.4}) takes form

\begin{eqnarray}
F(r)=1-\frac{2M}{r}+\frac{Q^{2}}{r^{2}}+r^{2}-\frac{aQ^{4}}{20r^{6}},
\label{eq3.1}
\end{eqnarray}

\noindent and then the surface gravity is $\kappa=\frac{1}{2r_+}\left(1 - \frac{Q^2}{r_+^2}+3 r_+^2+\frac{aQ^4}{4r_+^6}\right)$. When $\Lambda=0$, Eq. (\ref{eq2.4}) becomes

\begin{eqnarray}
F(r)=1-\frac{2M}{r}+\frac{Q^{2}}{r^{2}}-\frac{aQ^{4}}{20r^{6}},
\label{eq3.2}
\end{eqnarray}

\noindent and the surface gravity is $\kappa=\frac{1}{2r_+}\left(1 - \frac{Q^2}{r_+^2} +\frac{aQ^4}{4r_+^6}\right)$. Inserting Eqs. (\ref{eq3.1}) and (\ref{eq3.2}) into Eq. (\ref{eq2.11}) and using Eq. (\ref{eq2.13}), we get the expressions of the orbits. Due to their complexities, we let the parameters of the black hole and particle be certain values to obtain the positions $r_0$ of some certain orbits. 

\subsection{Fixing particle's charge and changing its angular momentum}

In this subsection, we order $M = 1$ and $ q = 15$. The positions $r_0$ of the circular orbits are gotten by changing the angular momentum and listed in Table 1-Table 5. The horizon radius is numerically gotten. 

\begin{table}[H]
	\begin{center}
		\setlength{\tabcolsep}{2.5mm}
		\begin{tabular}{ccccccccc}
			\toprule[1pt]
			&  L &  0 &  1 &  2 & 3 &  5 &  10 &  20 \\  \Xcline{2-9}{0.3pt}
			\multirow{5}*{$r_0$} & a=1.0  & 0.907044 & 0.920979 & 0.954082 & 0.994850	& 1.080621 & 1.273581 & 1.548244  \\  \Xcline{2-9}{0.3pt}
			& a=0.8  & 0.904490	& 0.916975 & 0.947887 & 0.987382 & 1.072733	& 1.267649 & 1.544991  \\   \Xcline{2-9}{0.3pt}
			& a=0.6  & 0.902236	& 0.913399 & 0.942085 & 0.980113 & 1.064787	& 1.261615 & 1.541707  \\   \Xcline{2-9}{0.3pt}
			& a=0.3  & 0.899289	& 0.908722 & 0.934129 & 0.969656 & 1.052787	& 1.252361 & 1.536719  \\   \Xcline{2-9}{0.3pt}
			& a=0  & 0.896733 & 0.904721 & 0.927047	& 0.959819 & 1.040741 & 1.242852 & 1.531654  \\   
			\bottomrule[1pt]
		\end{tabular}
		\label{tab1}
	\end{center}
	Table 1. Positions of circular orbits of a charged particle around the EEH AdS black hole, where $\Lambda = -3$ and $Q=0.6$. The event horizon is located at $r_+=0.894048$ when $a=1.0$, at $r_+=0.893291$ when $a=0.8$, at $r_+=0.892526$ when $a=0.6$, at $r_+=0.891364$ when $a=0.3$ and at $r_+=0.890184$ when $a=0$. 
\end{table}

\begin{table}[H]
	\begin{center}
		\setlength{\tabcolsep}{2.5mm}
		\begin{tabular}{ccccccccc}
			\toprule[1pt]
			&  L &  0 &  1 &  2 & 3 &  5 &  10 &  20 \\  \Xcline{2-9}{0.3pt}
			\multirow{5}*{$r_0$} & a=1.0  & 0.857517 & 0.871160 & 0.900337 & 0.933774	& 1.001922 & 1.159156 & 1.402398  \\   \Xcline{2-9}{0.3pt}
			& a=0.8  & 0.851938 & 0.863014 & 0.888933 & 0.920650 & 0.988113 & 1.147949 & 1.395847  \\   \Xcline{2-9}{0.3pt}
			& a=0.6  & 0.847348	& 0.856198 & 0.878595 & 0.908001 & 0.974043	& 1.136302 & 1.389141  \\   \Xcline{2-9}{0.3pt}
			& a=0.3  & 0.841572	& 0.847857 & 0.865193 & 0.890328 & 0.952625	& 1.117915 & 1.378767  \\   \Xcline{2-9}{0.3pt}
			& a=0  & 0.836492 & 0.841000 & 0.854103	& 0.874625 & 0.931193 & 1.098282 & 1.367983  \\   
			\bottomrule[1pt]
		\end{tabular}
		\label{tab1}
	\end{center}
	Table 2. Positions of circular orbits of a charged particle around the EEH AdS black hole, where $\Lambda = -3$ and $Q=0.7$. The event horizon is located at $r_+=0.844860$ when $a=1.0$, at $r_+=0.842699$ when $a=0.8$, at $r_+=0.840469$ when $a=0.6$, at $r_+=0.836986$ when $a=0.3$ and at $r_+=0.833318$ when $a=0$.
\end{table}

\begin{table}[H]
	\begin{center}
		\setlength{\tabcolsep}{2.5mm}
		\begin{tabular}{ccccccccc}
			\toprule[1pt]
			&  L &  0 &  1 &  2 & 3 &  5 &  10 &  20 \\  \Xcline{2-9}{0.3pt}
			\multirow{5}*{$r_0$} & a=1.0  & 1.447628 & 1.449130 & 1.453559 & 1.460695 & 1.481730 & 1.556220 &  1.707025  \\   \Xcline{2-9}{0.3pt}
			& a=0.8  & 1.445805 & 1.447240 & 1.451475 & 1.458321 & 1.478635 & 1.551775 & 1.702627  \\   \Xcline{2-9}{0.3pt}
			& a=0.6  & 1.443969	& 1.445339 & 1.449389 & 1.455954 & 1.475562	& 1.547319 & 1.698174  \\   \Xcline{2-9}{0.3pt}
			& a=0.3  & 1.441187	& 1.442465 & 1.446252 & 1.452416 & 1.470992	& 1.540618 & 1.691391  \\   \Xcline{2-9}{0.3pt}
			& a=0  & 1.438368 & 1.439561 & 1.443102 & 1.448886 & 1.466463 & 1.533900 & 1.684475  \\   
			\bottomrule[1pt]
		\end{tabular}
		\label{tab1}
	\end{center}
	Table 3. Positions of circular orbits of a charged particle around the EEH black hole, where $Q=0.9$. The event horizon is located at $r_+=1.444450$ when $a=1.0$, at $r_+=1.442783$ when $a=0.8$, at $r_+=1.441094$ when $a=0.6$, at $r_+=1.438518$ when $a=0.3$ and at $r_+=1.435890$ when $a=0$.
\end{table}

\begin{table}[H]
	\begin{center}
		\setlength{\tabcolsep}{2.5mm}
		\begin{tabular}{ccccccccc}
			\toprule[1pt]
			&  L &  0 &  1 &  2 & 3 &  5 &  10 &  20 \\  \Xcline{2-9}{0.3pt}
			\multirow{5}*{$r_0$} & a=1.0  & 1.334404 & 1.335576 & 1.339041 & 1.344660 & 1.361467 &  1.423745 & 1.561036  \\   \Xcline{2-9}{0.3pt}
			& a=0.8  & 1.330490 & 1.331560 & 1.334770 & 1.339985 & 1.355750 & 1.415822 & 1.552969  \\   \Xcline{2-9}{0.3pt}
			& a=0.6  & 1.326463 & 1.327457 & 1.330410 & 1.335242 & 1.349995  & 1.407772 & 1.544650  \\   \Xcline{2-9}{0.3pt}
			& a=0.3  & 1.320185 & 1.321061 & 1.323673 & 1.327969 & 1.341261 & 1.395443 & 1.531650  \\   \Xcline{2-9}{0.3pt}
			& a=0  & 1.313580 & 1.314350 & 1.316651 & 1.320454 & 1.332360 & 1.382786 & 1.517948  \\   
			\bottomrule[1pt]
		\end{tabular}
		\label{tab1}
	\end{center}
	Table 4. Positions of circular orbits of a charged particle around the EEH black hole, where $Q=0.95$. The event horizon is located at $r_+=1.332304$ when $a=1.0$, at $r_+=1.328569$ when $a=0.8$, at $r_+=1.324707$ when $a=0.6$, at $r_+=1.318655$ when $a=0.3$ and at $r_+=1.312250$ when $a=0$. 
\end{table}

\begin{table}[H]
	\begin{center}
		\setlength{\tabcolsep}{2.5mm}
		\begin{tabular}{ccccccccc}
			\toprule[1pt]
			&  L &  0 &  1 &  2 & 3 &  5 &  10 &  20 \\  \Xcline{2-9}{0.3pt}
			\multirow{4}*{$r_0$} & Q=0.76  & 0.784386 & 0.787263 & 0.795811 & 0.809757 & 0.851533 & 0.995822 & 1.256175  \\   \Xcline{2-9}{0.3pt}
			& Q=0.74  & 0.803891 & 0.807291	& 0.817319 & 0.833445 & 0.880289 & 1.032450 & 1.295411  \\   \Xcline{2-9}{0.3pt}
			& Q=0.72  & 0.821077 & 0.825019	& 0.836556 & 0.854864 & 0.906611 & 1.066345 & 1.332510  \\   \Xcline{2-9}{0.3pt}
			& Q=0.70  & 0.836492 & 0.841000	& 0.854103 & 0.874625 & 0.931193 & 1.098282	& 1.367983  \\      
			\bottomrule[1pt]
		\end{tabular}
		\label{tab1}
	\end{center}
	Table 5. Positions of circular orbits of a charged particle around the RN AdS black hole, where $\Lambda = -3$. The event horizon is located at $r_+=0.782618$ when $Q=0.76$, at $r_+=0.801691$ when $Q=0.74$, at $r_+=0.818410$ when $Q=0.72$ and at $r_+=0.833318$ when $Q=0.70$.
\end{table}

From the tables, it is found that when the angular momentum of the particle increases, the positions of the orbits gradually move away from the horizon, but they finally tend to the certain positions. For the fixed angular momenta and charges, the orbits gradually move away from the horizon with the increase of the value of the EH parameter. In Table 1, there is $r_0 = 1.0145r_+$ for $a=1$ and $r_0 = 1.0088r_+$ for $a=0.3$ when $L=0$, which shows that the orbit is close to the horizon of the EEH AdS black hole when the angular momentum is zero and the value of $a$ is small enough.  

\begin{figure}[htb]
	\centering
	\includegraphics[width=12cm,height=8cm]{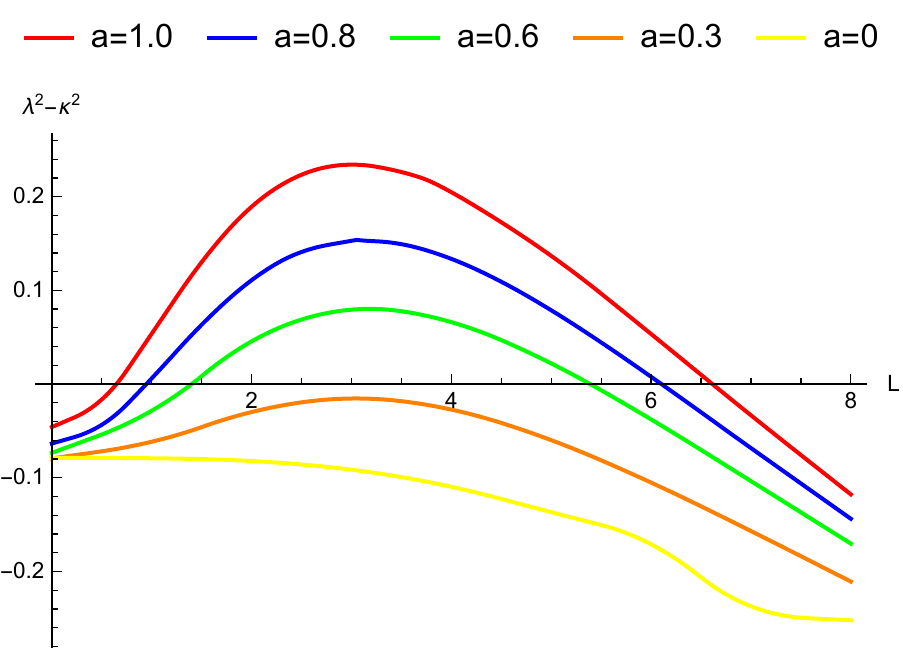}
	\label{Figure 1}
	\caption{The influence of the angular momentum of the charged particle around the EEH AdS black hole on the Lyapunov exponent, where $\Lambda = -3$ and $Q=0.6$. The violation for the chaos bound occurs at $0.63<L<6.62$ (The corresponding spatial region is $1.0210r_+<r_0<1.2836r_+$.) when $a= 1$, at $0.94<L<6.12$ ($1.0250r_+<r_0<1.2535r_+$) when $a= 0.8$ and at $1.36<L<5.38$ ($1.0332r_+<r_0<1.2111r_+$) when $a= 0.6$.}
\end{figure}

\begin{figure}[htb]
	\centering
	\includegraphics[width=12cm,height=8cm]{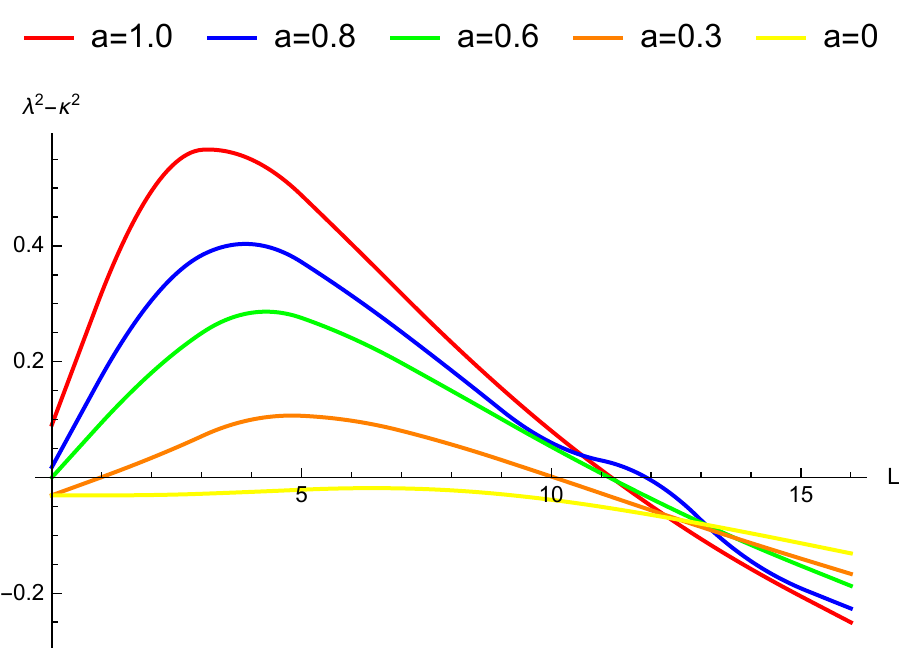}
	\caption{The influence of the angular momentum of the charged particle around the EEH AdS black hole on the Lyapunov exponent, where $\Lambda = -3$ and $Q=0.7$. The violation for the chaos bound occurs at $0\leq L<11.20$ $( 1.0150r_+ \leq r_0<1.4123r_+)$ when $a= 1$, at $0\leq L<11.05$ $(1.0110r_+ \leq r_0 <1.3985r_+)$ when $a= 0.8$, at $0 \leq L<11.15$ $(1.0082r_+ \leq r_0 <1.3926r_+)$ when $a= 0.6$ and at $0\leq L<10.05$ $1.0055r_+ \leq r_0 <1.3375r_+)$ when $a= 0.3$. }
\end{figure}

\begin{figure}[htb]
	\centering
	\includegraphics[width=12cm,height=8cm]{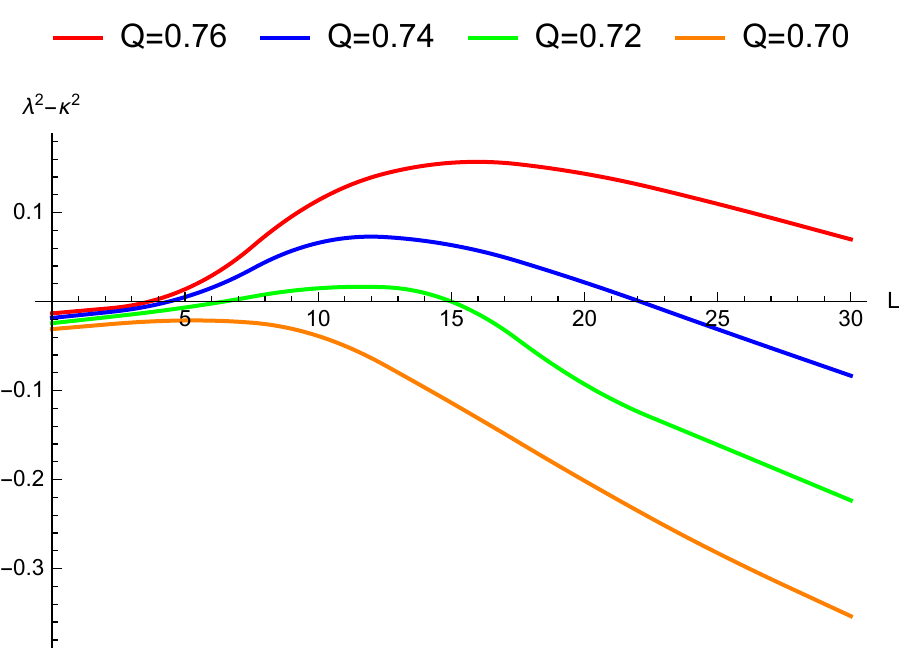}
	\caption{The influence of the angular momentum of the charged particle around the RN AdS black hole on the Lyapunov exponent, where $\Lambda = -3$. There are violations for the chaos bound when $Q=0.72, 0.74, 0.76$. The violation occurs at $4.01<L<39.08$ $(1.0590r_+<r_0<1.9956r_+)$ when $Q=0.76$, at $4.59<L<22.09$$(1.0845r_+<r_0<1.6703r_+)$ when $Q=0.74$ and at $5.76<L<15.00$ $(1.1360r_+<r_0<1.4811r_+)$ when $Q=0.72$.}
\end{figure}

\begin{figure}[htb]
	\centering
	\includegraphics[width=12cm,height=8cm]{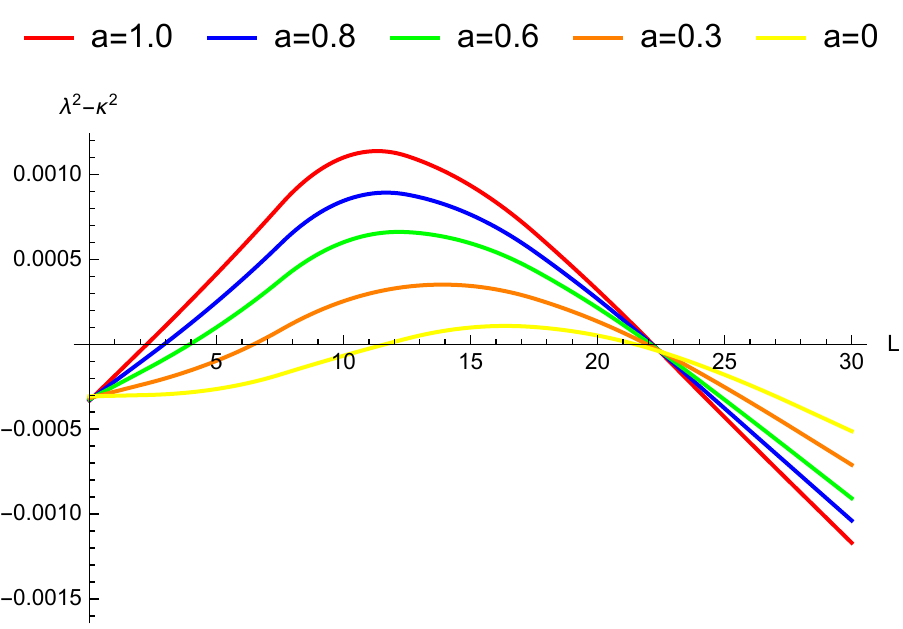}
	\caption{The influence of the angular momentum of the charged particle around the EEH black hole on the Lyapunov exponent, where $Q=0.9$. There are violations for the chaos bound when $a=0, 0.3, 0.6, 0.8, 1.0$. The violation occurs at $3.03 <L< 22.16$ $(1.0114r_+ <r_0 <1.2007r_+)$ when $a=1$, at $3.52< L<22.13$ $(1.0139r_+ < r_0 <1.1989r_+)$ when $a= 0.8$, at $4.24< L<22.07$ $(1.0181r_+< r_0 <1.1969r_+)$ when $a= 0.6$, at $6.28< L<21.91$ $(1.0333r_+ < r_0 <1.1931r_+)$ when $a= 0.3$ and at $11.55<L<21.42$ $(1.0851r_+ < r_0 <1.1862r_+ )$ when $a=0$.}
\end{figure}

\begin{figure}[htb]
	\centering
	\includegraphics[width=12cm,height=8cm]{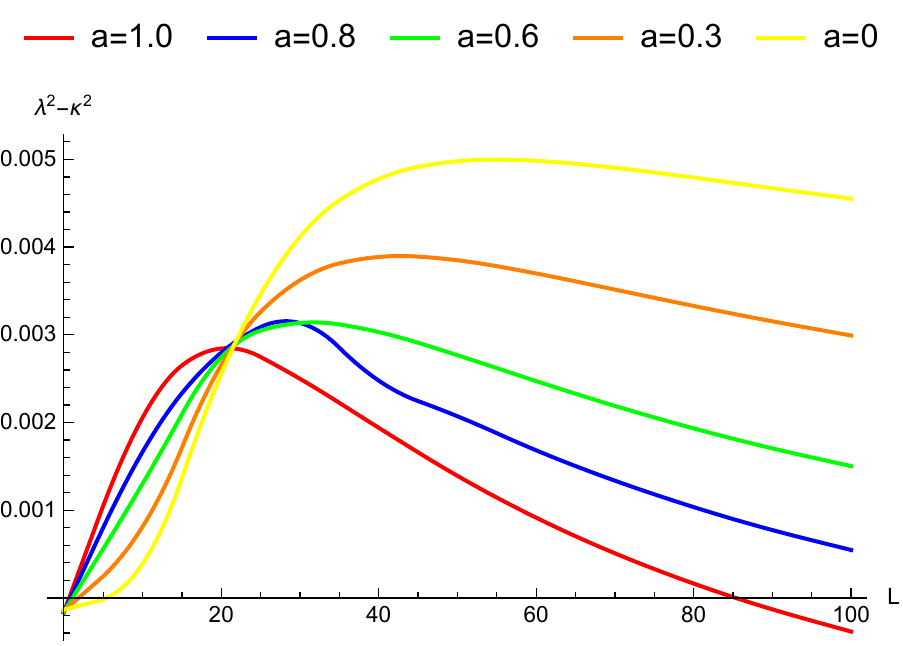}
	\caption{The influence of the angular momentum of the charged particle around the EEH black hole on the Lyapunov exponent, where $Q=0.95$. There are violations for the chaos bound when $a=0, 0.3, 0.6, 0.8, 1.0$. When $a=1$, the violation occurs at $1.93< L< 85.40$ $(1.0048r_+<r_0<1.4486r_+)$.}
\end{figure}

The values of the Lyapunov exponent at the orbits $r=r_0$ and the surface gravity are  numerically calculated by using Eqs. (\ref{eq2.5}), (\ref{eq2.14}), (\ref{eq3.1}) and (\ref{eq3.2}). The violation of the chaos bound in specific regions is analyzed in Figure 1-Figure 5. For the EEH AdS black hole with a fixed charge, there are violations for the chaos bound when the value of $a$ is relatively large. In Figure 1, the range of the angular momentum and spatial region where the chaos bound is violated increase with the increase of the value of $a$. In Figure 2, the violation occurs when $a=1$, $0\leq L <11.20$ $( 1.0150r_+ \leq r_0<1.4123r_+)$, and when $a = 0.8$, $0\leq L <11.05$ $(1.0110r_+ \leq r_0 <1.3985r_+)$. $L = 0$ means that the particle is stationary relative to the black hole. This implies that as long as the values of the charge and EH parameter are large enough, the angular momentum is small enough, or even is zero, the bound can be violated. However, when the angular momentum is large, there is no violation. Therefore, the spatial region for the violation is affected by the EH parameter, charge and angular momentum. When $a = 0.3$, the violation occurs when $Q=0.7$, and does not occur when $Q=0.6$. This shows that the large charge of a black hole is more likely to cause the violation. When $a=0$, the metric (\ref{eq2.2}) describes the RN AdS black hole and there is no violation for the bound in Figure 1 and Figure 2. However, it does not imply that there is no violation for the chaos bound of a charged particle around the black hole. This is because the value of the black hole's charge is relatively small. When the charge is large enough, the violation is found in Figure 3. From the figure, it is shown again that the spatial region where the bound is violated is large when the charge of the black hole is large.

When $\Lambda =0$, the chaos bound of the particle around the EEH black hole with a fixed charge is described in Figure 4 and Figure 5. From the figures, it is found that the bound is violated for all the values of $a$ when the charge is large. For the fixed EH parameter, the violation occurs in a large range of the angular momentum (spatial region) when the charge is relatively large. When the charge takes the different values, the changes of the values of $\lambda^2-\kappa^2$ are significantly different. For the fixed charge and large value of the EH parameter, the values of $\lambda^2-\kappa^2$ increase faster when the angular momentum is less than $20$, and decrease faster when the angular momentum is greater than $20$. This is more evident in Figure 5. However, the angular momentum region and spatial region where the bound is violated decrease with the increase of the value of $a$ in Figure 5. This is different from the first four pictures.

Therefore, for the certain values of the EH parameter, charge and cosmological constant, the chaos bound is violated in the near-horizon region and at a certain distance from the horizon. The angular momentum of the particle plays an important role in the violation of the bound. Because other parameters are fixed, the value of the angular momentum determines the positions of the orbits and the region where the bound is violated. When the value of the angular momentum is small, the orbit is very close to the horizon. With the increase of the value of the angular momentum, the orbits gradually move away from the horizon and tend to constant positions.

\subsection{Fixing particle's angular momentum and changing its charge}

In this subsection, we order $M = 1$ and $L = 5$. The positions $r_0$ of the circular orbits are derived by changing the charge of the particle and listed in Table 6-Table 9. In the tables, "$\ast\ast$$\ast\ast $" indicates that circular orbits don't exist. The locations of the horizons are also obtained by numerical calculations. A common feature in these tables is that when the particle's charge increases, the orbits gradually approach the horizons. In Table 6, there is no circular orbit around the EEH AdS black hole when $q\leq 3$. When $a=0$, the EEH AdS black hole is reduced to the RN AdS black hole. There is also no circular orbit for $q\leq 2$ in Table 7. When $\Lambda = 0$, circular orbits around the EEH black hole are found for $q>1$ and $q \leq 1$ in Table 8 and 9.

\begin{table}[H]
	\begin{center}
		\setlength{\tabcolsep}{2.5mm}
		\begin{tabular}{ccccccccc}
			\toprule[1pt]
			&  $q$ &  0.5 &  1 &  2 & 3 & 4 &  5 &  10 \\  \Xcline{2-9}{0.3pt}
			\multirow{5}*{$r_0$} & a=1.0  & **** & **** & ****  & **** & 1.499239 &	1.355364 &	1.091599 \\  \Xcline{2-9}{0.3pt}
			& a=0.8  & **** & **** & ****  & **** & 1.490236 &	1.345668 &	1.078504  \\   \Xcline{2-9}{0.3pt}
			& a=0.6  & **** & **** & ****  & **** & 1.481070 &	1.335704 &	1.064915  \\   \Xcline{2-9}{0.3pt}
			& a=0.3  & **** & **** & ****  & **** & 1.466991 &	1.320205 &	1.043546  \\   \Xcline{2-9}{0.3pt}
			& a=0  & **** & **** & ****  & **** & 1.452472 &	1.303962 &	1.020926 \\   
			\bottomrule[1pt]
		\end{tabular}
		\label{tab1}
	\end{center}
	Table 6. Positions of circular orbits of a particle around the EEH AdS black hole, where $\Lambda = -3$ and $Q=0.7$. The event horizon is located at $r_+=0.844860$ when $a=1.0$, at $r_+=0.842699$ when $a=0.8$, at $r_+=0.840469$ when $a=0.6$, at $r_+=0.836986$ when $a=0.3$ and at $r_+=0.833318$ when $a=0$.
\end{table}

\begin{table}[H]
	\begin{center}
		\setlength{\tabcolsep}{2.5mm}
		\begin{tabular}{ccccccccc}
			\toprule[1pt]
			&  $q$ &  0.5 &  1 &  2 & 3 &  4 &  5 & 10 \\  \Xcline{2-9}{0.3pt}
			\multirow{5}*{$r_0$} & Q=0.76  & **** & **** & **** & 1.537130 &	1.299666 &	1.174371 &	0.924467  \\  \Xcline{2-9}{0.3pt}
			& Q=0.74  & **** & **** & ****  & 1.621182 &	1.350527 &	1.219018 &	0.958992  \\   \Xcline{2-9}{0.3pt}
			& Q=0.72 & **** & **** & **** & 1.734581 &	1.401095 &	1.261953 &	0.990881  \\   \Xcline{2-9}{0.3pt}
			& Q=0.70  & **** & **** & ****  & **** & 1.452472 &	1.303962 &	1.020926  \\   
			\bottomrule[1pt]
		\end{tabular}
		\label{tab1}
	\end{center}
	Table 7. Positions of circular orbits of a particle around the RN AdS black hole, where $\Lambda = -3$. The event horizon is located at $r_+=0.782618$ when $Q=0.76$, at $r_+=0.801691$ when $Q=0.74$, at $r_+=0.818410$ when $Q=0.72$ and at $r_+=0.833318$ when $Q=0.70$.
\end{table}

\begin{table}[H]
	\begin{center}
		\setlength{\tabcolsep}{2.5mm}
		\begin{tabular}{ccccccccc}
			\toprule[1pt]
			&  $q$ &  0.5 &  1 &  2 & 3 &  4 &  5 & 10 \\  \Xcline{2-9}{0.3pt}
			\multirow{5}*{$r_0$} & a=1.0  & 2.011500 &	1.842216 &	1.518311 &	1.485960 &	1.404786 &	1.352891 &	1.254538   \\   \Xcline{2-9}{0.3pt}
			& a=0.8  & 2.007721 &	1.835878 &	1.606614 &	1.470208 &	1.386966 &	1.334511 &	1.239276  \\   \Xcline{2-9}{0.3pt}
			& a=0.6  & 2.003875 &	1.829340 &	1.594156 &	1.453004 &	1.367308 &	1.314264 &	1.222888  \\   \Xcline{2-9}{0.3pt}
			& a=0.3  & 1.997973 &	1.819123 &	1.573722 &	1.423532 &	1.332973 & 1.278960 &	1.195276  \\   \Xcline{2-9}{0.3pt}
			& a=0  & 1.991903 &	1.808352 &	1.550538 &	1.387280 &	1.288877 &	1.233642 &	1.161352  \\   
			\bottomrule[1pt]
		\end{tabular}
		\label{tab1}
	\end{center}
	Table 8. Positions of circular orbits of a particle around the EEH black hole, where $Q=0.99$. The event horizon is located at $r_+=1.206347$ when $a=1.0$, at $r_+=1.196578$ when $a=0.8$, at $r_+=1.185690$ when $a=0.6$, at $r_+=1.166388$ when $a=0.3$ and at $r_+=1.141067$ when $a=0$. 
\end{table}

\begin{table}[H]
	\begin{center}
		\setlength{\tabcolsep}{2.5mm}
		\begin{tabular}{ccccccccc}
			\toprule[1pt]
			&  $q$ &  0.5 &  1 &  2 & 3 &  4 &  5 & 10 \\  \Xcline{2-9}{0.3pt}
			\multirow{5}*{$r_0$} & a=1.0  & 2.166321 &	2.003486 &	1.785054 &	1.651170 &	1.565267 &	1.507993 &	1.392599   \\   \Xcline{2-9}{0.3pt}
			& a=0.8  & 2.164209 &	2.000086 &	1.779213 &	1.643572 &	1.556737 &	1.499199 &	1.385459  \\   \Xcline{2-9}{0.3pt}
			& a=0.6  & 2.162080 &	1.996641 &	1.773234 &	1.635746 &	1.547954 &	1.409159 &	1.378258  \\   \Xcline{2-9}{0.3pt}
			& a=0.3  & 2.158853 &	1.991383 &	1.763986 &	1.623539 &	1.534233 &	1.476095 &	1.367328  \\   \Xcline{2-9}{0.3pt}
			& a=0  & 2.155584 &	1.986014 &	1.754373 &	1.610700 &	1.519779 &	1.461362 &	1.356214  \\   
			\bottomrule[1pt]
		\end{tabular}
		\label{tab1}
	\end{center}
	Table 9. Positions of circular orbits of a particle around the EEH black hole, where $Q=0.95$. The event horizon is located at $r_+=1.332304$ when $a=1.0$, at $r_+=1.328569$ when $a=0.8$, at $r_+=1.324707$ when $a=0.6$, at $r_+=1.318655$ when $a=0.3$ and at $r_+=1.312250$ when $a=0$. 
\end{table}

\begin{figure}[htb]
	\centering
	\includegraphics[width=12cm,height=8cm]{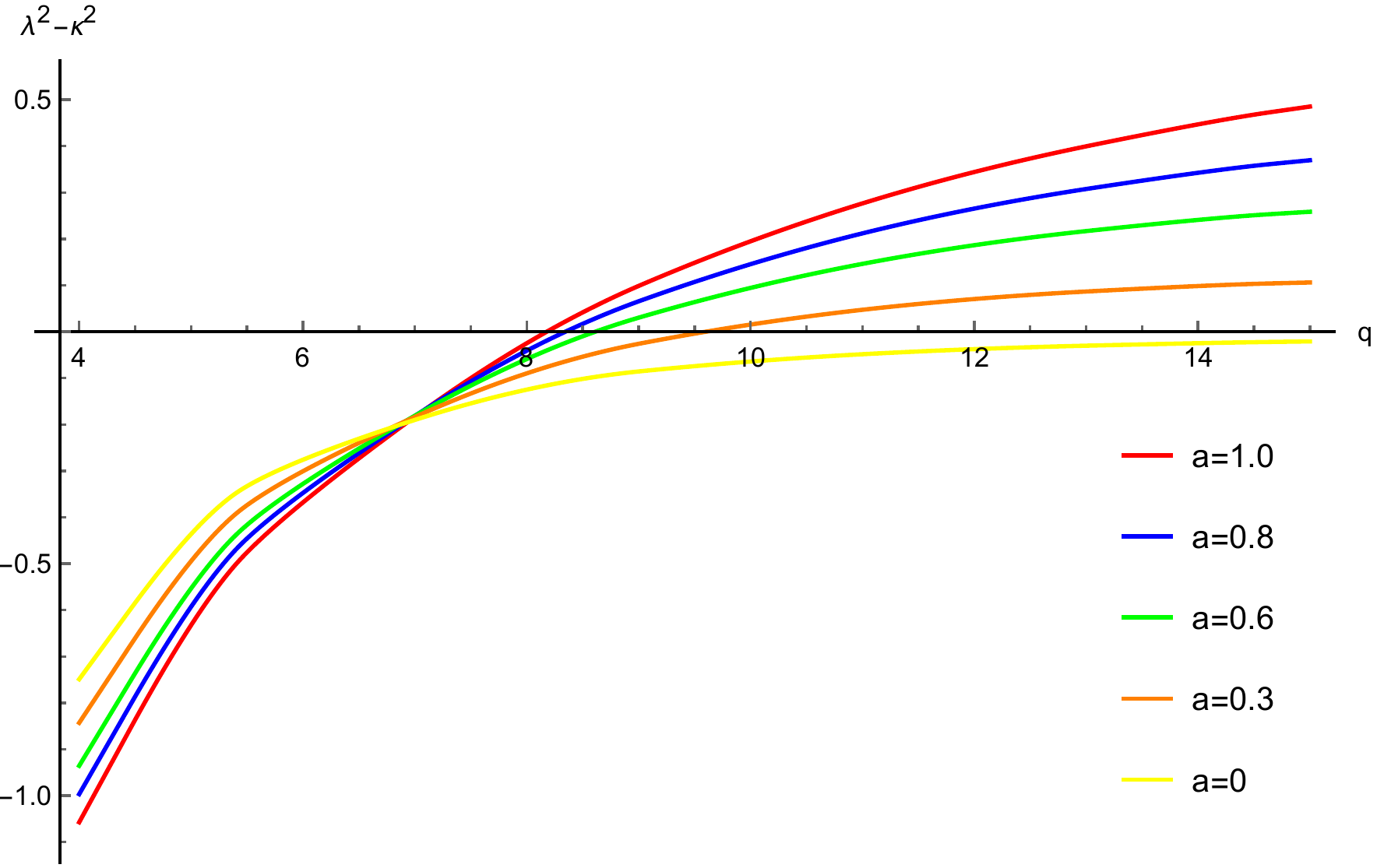}
	\caption{The influence of the charge of the particle around the EEH AdS black hole on the Lyapunov exponent, where $\Lambda = -3$ and $Q=0.7$.}
\end{figure}

\begin{figure}[htb]
	\centering
	\includegraphics[width=12cm,height=8cm]{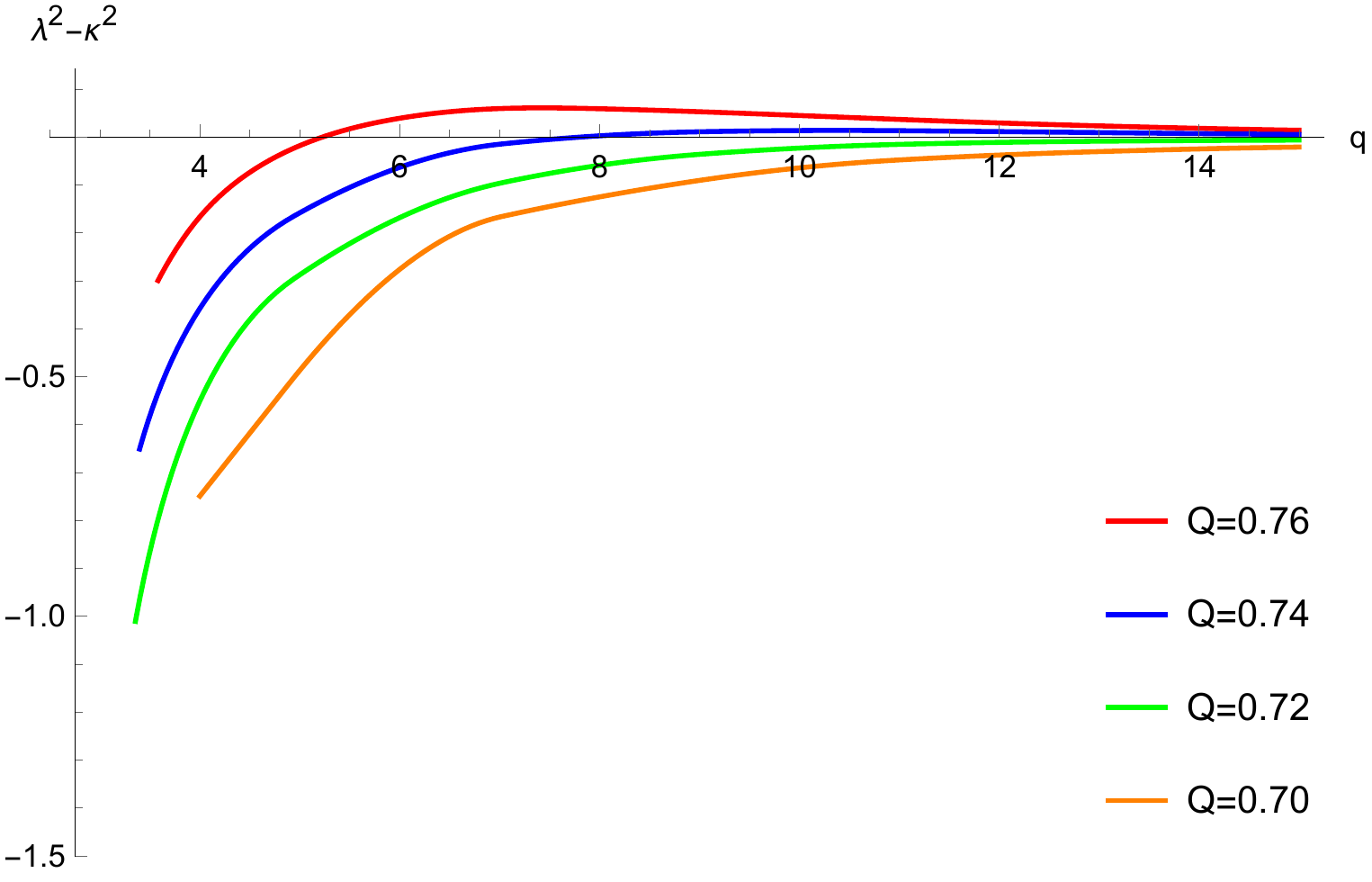}
	\caption{The influence of the charge of the particle around the RN AdS black hole on the Lyapunov exponent, where $\Lambda = -3$.}
\end{figure}

\begin{figure}[htb]
	\centering
	\includegraphics[width=12cm,height=8cm]{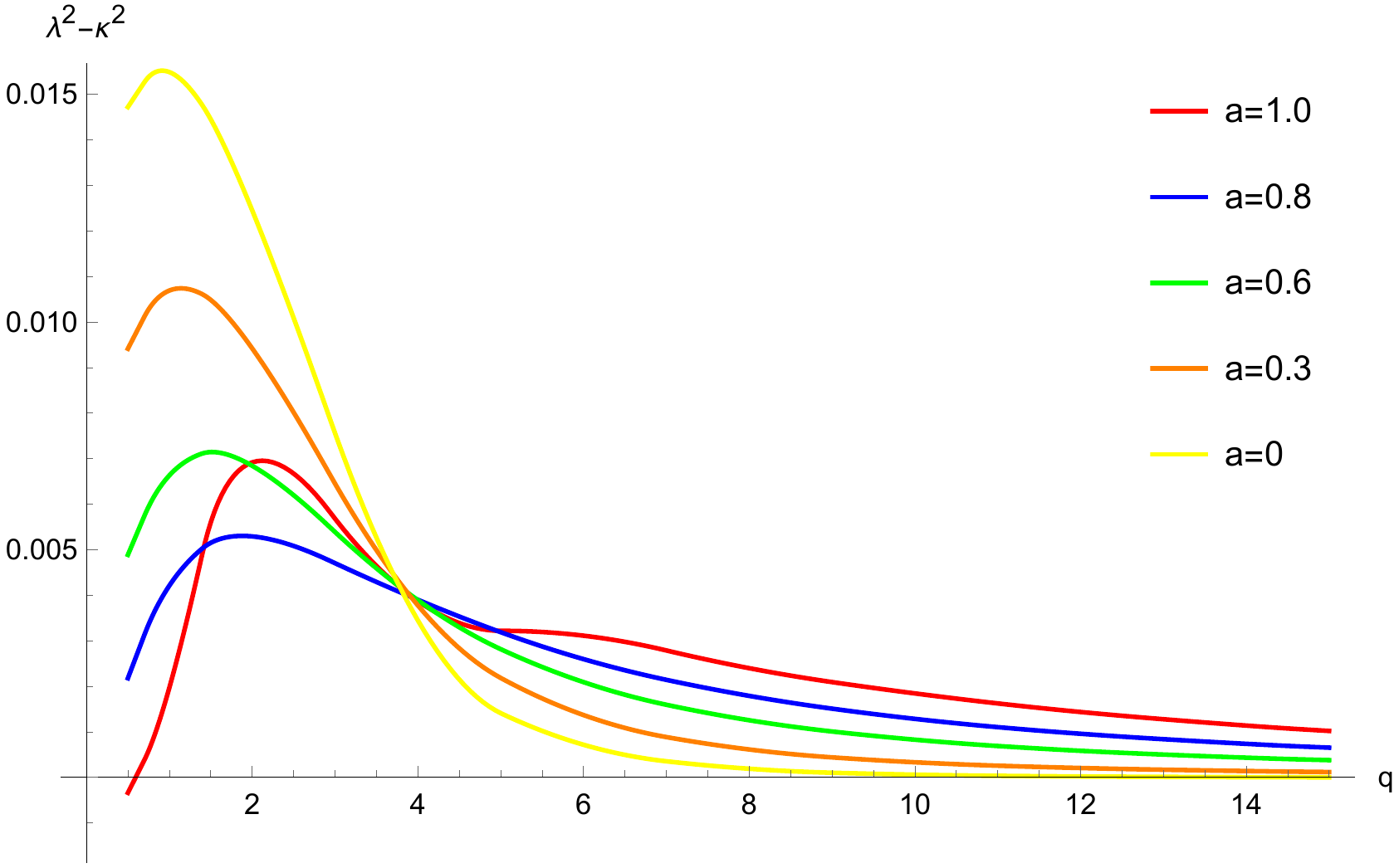}
	\caption{The influence of the charge of the particle around the EEH black hole on the Lyapunov exponent, where $Q=0.99$.}
\end{figure}

\begin{figure}[htb]
	\centering
	\includegraphics[width=12cm,height=8cm]{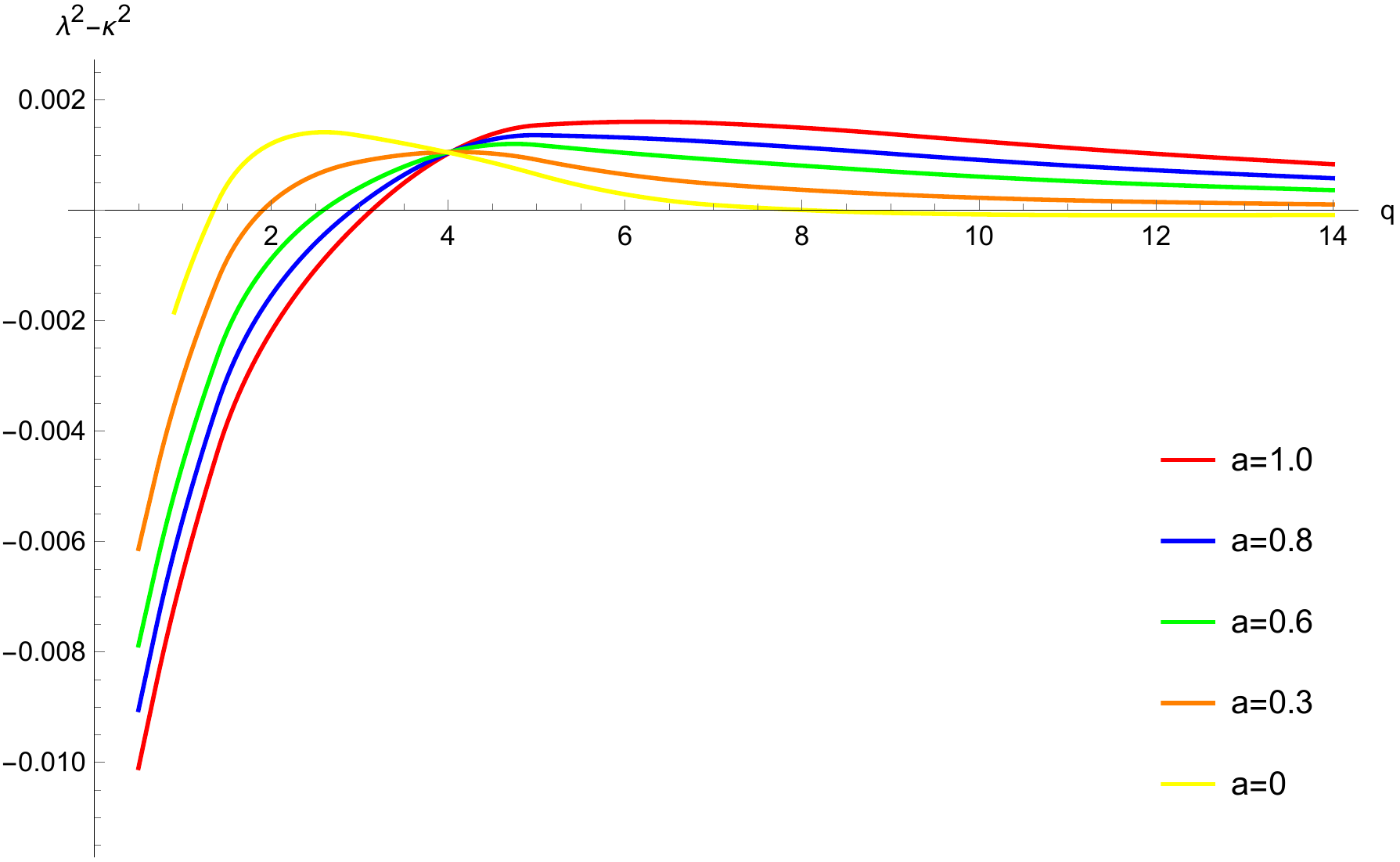}
	\caption{The influence of the charge of the particle around the EEH black hole on the Lyapunov exponent, where $Q=0.95$.}
\end{figure}

The values of the exponent and surface gravity are also numerically calculated by using Eqs. (\ref{eq2.5}), (\ref{eq2.14}), (\ref{eq3.1}) and (\ref{eq3.2}), and their relations are plotted in Figure 6-Figure 9. In Figure 6, the values of $\lambda^2- \kappa^2$ increase with the increase of the particle's charge. When the charge increases to a certain value, the violation occurs. Its increase speed is related to the EH parameter. The larger the parameter, the greater the increase speed. For $L=5$, the large parameter is more likely to violate the bound. When $a = 0$, there is no violation in the figure. However, this does not mean that the bound isn't violated. The reason for this is that the black hole's charge is not large enough. When this charge is large enough, the violation appears. For convenience of comparison, we let $a = 0$ and plot a picture in Figure 7. This figure shows that a large charge of the black hole is more likely to cause the violation. When the cosmological constant is 0, the violation in the EEH spacetime is described in Figure 8 and 9. Comparing Figure 8 and 9, we find that when the black hole's charge is large enough (for example, $Q = 0.99$ ), the violation appears for $q<1$. However, this phenomenon does not exist when $Q = 0.95$. Figure 8 and 9 reflects that the range of the value of $q$ corresponding to the violation increases with the increase of $a$.

\section{Conclusions and discussions}

In this paper, we investigated the influences of the angular momentum and charge of the particle around the EEH AdS black hole on the Lyapunov exponent. For the specific parameters of the black hole, the spatial regions where the chaos bound is violated were found by fixing the particle's charge and changing its angular momentum. 

When the charge is fixed at $q=15$, the positions of the circular orbits were gotten by changing the angular momentum. These positions gradually move away from the horizons with the increase of the angular momentum. The spatial region and angular momentum's range where the violation occurs increase with the increase of the values of the EH parameter. For the fixed cosmological constant and EH parameter, the large charge of the black hole is more likely to cause the violation. When the value of the EH parameter is large, a small angular momentum causes the violation. Although the angular momentum's value is small, its corresponding spatial region is not small. For example, when $\Lambda = -3$, $Q = 0.7$ and $a = 1$, the violation occurs in the range $0\leq L<11.20$. The corresponding spatial region is $1.0150r_+ \leq r_0<1.4123r_+$. When $\Lambda = 0$, $Q = 0.95$ and $a = 1$, the violation occurs in the range $1.93< L< 85.40$ and the spatial region is $1.0048r_+<r_0<1.4486r_+$. When the angular momentum is fixed at $L = 5$, the circular orbits were obtained by changing the particle's charge. As the charge increases, the positions gradually approach the horizons. The violation of the bound was found. For $\Lambda =- 3$, the bound is violated when the particle's charge is relatively large. An interesting discovery is that the bound is violated by the black hole when the particle’s charge is less than 1  and $\Lambda =0$, but this requires the black hole's charge to be large enough. In  \cite{KG1,KG2}, the violation was found when the particle’s mass is 1 and its charge is 10. The weak gravity conjecture asserts that, for the lightest charged particle along the direction of a basis vector in charge space, the charge-to-mass ratio is larger than those for extremal black holes. For an extremal Kerr-Newman black hole, when its rotation parameter is very small, the charge-to-mass ratio of the black hole approaches 1. Therefore,  the bound can be violated without the premise of the weak gravity conjecture. 

There are two explanations for the violation of the bound. In \cite{LTW1,LTW2}, the authors discussed the minimal length effects on the chaotic motion of the particle, and found that the bound is violated. They believed that the result does not necessarily implies the violation for the bound conjectured in \cite{MSS}, and the bound should be corrected due to the minimal length in bulk. On the other hand, Lei and Ge found that the bound is violated by the RN and RN AdS black holes \cite{LG2}. They perceived that this violation should be affected by some properties of the black hole and related to the dynamical stability of the black hole.  

Although we found that the bound is violated, the backreaction of the particle on the background spacetime wasn't considered. When it is taken into account, the value of the exponent should be changed, and its relation with the bound needs to be further studied. This is our next work and may lead to an interesting result. If the bound is still violated, we will understand and study this violation from the stability of black holes. The study on the stability of the black holes may help us to understand the exceeding of the bound.

\acknowledgments
We are grateful to Haitang Yang, Peng Wang and Jun Tao for useful discussions. This work is supported by the NSFC (Grant No. 12105031) and Tianfu talent project.

\end{document}